\documentstyle[epsf,12pt]{article}

\catcode`\@=11
\long\def\@makefntext#1{
\protect\noindent \hbox to 3.2pt {\hskip-.9pt  
$^{{\ninerm\@thefnmark}}$\hfil}#1\hfill}                

\def\@makefnmark{\hbox to 0pt{$^{\@thefnmark}$\hss}}  
        
\def\ps@myheadings{\let\@mkboth\@gobbletwo
\def\@oddhead{\hbox{}
\rightmark\hfil\ninerm\thepage}   
\def\@oddfoot{}\def\@evenhead{\ninerm\thepage\hfil
\leftmark\hbox{}}\def\@evenfoot{}
\def\sectionmark##1{}\def\subsectionmark##1{}}

\renewcommand{\thefootnote}{\fnsymbol{footnote}}

\newcounter{sectionc}\newcounter{subsectionc}\newcounter{subsubsectionc}
\renewcommand{\section}[1] {\vspace*{0.6cm}\addtocounter{sectionc}{1} 
\setcounter{subsectionc}{0}\setcounter{subsubsectionc}{0}\noindent 
        {\normalsize\bf\thesectionc. #1}\par\vspace*{0.4cm}}
\renewcommand{\subsection}[1] {\vspace*{0.6cm}\addtocounter{subsectionc}{1} 
        \setcounter{subsubsectionc}{0}\noindent 
        {\normalsize\it\thesectionc.\thesubsectionc. #1}\par\vspace*{0.4cm}}
\renewcommand{\subsubsection}[1]
{\vspace*{0.6cm}\addtocounter{subsubsectionc}{1}
        \noindent {\normalsize\rm\thesectionc.\thesubsectionc.\thesubsubsectionc. 
        #1}\par\vspace*{0.4cm}}

\newcounter{appendixc}
\newcounter{subappendixc}[appendixc]
\newcounter{subsubappendixc}[subappendixc]

\renewcommand{\appendix}[1] {\vspace*{0.6cm}
        \refstepcounter{appendixc}
        \setcounter{figure}{0}
        \setcounter{table}{0}
        \setcounter{equation}{0}
        \renewcommand{\thefigure}{\Alph{appendixc}.\arabic{figure}}
        \renewcommand{\thetable}{\Alph{appendixc}.\arabic{table}}
        \renewcommand{\theappendixc}{\Alph{appendixc}}
        \renewcommand{\theequation}{\Alph{appendixc}.\arabic{equation}}
        \noindent{\bf Appendix \theappendixc #1}\par\vspace*{0.4cm}}

\def\abstracts#1{{
        \centering{\begin{minipage}{12.2truecm}\footnotesize\baselineskip=12pt\noindent
        \centerline{\footnotesize ABSTRACT}\vspace*{0.3cm}
        \parindent=0pt #1
        \end{minipage}}\par}} 

\renewenvironment{thebibliography}[1]
        {\begin{list}{\arabic{enumi}.}
        {\usecounter{enumi}\setlength{\parsep}{0pt}
\setlength{\leftmargin 1.25cm}{\rightmargin 0pt}
         \setlength{\itemsep}{0pt} \settowidth
        {\labelwidth}{#1.}\sloppy}}{\end{list}}

\newcounter{itemlistc}
\newcounter{romanlistc}
\newcounter{alphlistc}
\newcounter{arabiclistc}

\newcommand{\fcaption}[1]{
        \refstepcounter{figure}
        \setbox\@tempboxa = \hbox{\footnotesize Fig.~\thefigure. #1}
        \ifdim \wd\@tempboxa > 6in
           {\begin{center}
        \parbox{6in}{\footnotesize\baselineskip=12pt Fig.~\thefigure. #1}
            \end{center}}
        \else
             {\begin{center}
             {\footnotesize Fig.~\thefigure. #1}
              \end{center}}
        \fi}

\newcommand{\tcaption}[1]{
        \refstepcounter{table}
        \setbox\@tempboxa = \hbox{\footnotesize Table~\thetable. #1}
        \ifdim \wd\@tempboxa > 6in
           {\begin{center}
        \parbox{6in}{\footnotesize\baselineskip=12pt Table~\thetable. #1}
            \end{center}}
        \else
             {\begin{center}
             {\footnotesize Table~\thetable. #1}
              \end{center}}
        \fi}

\def\@citex[#1]#2{\if@filesw\immediate\write\@auxout
        {\string\citation{#2}}\fi
\def\@citea{}\@cite{\@for\@citeb:=#2\do
        {\@citea\def\@citea{,}\@ifundefined
        {b@\@citeb}{{\bf ?}\@warning
        {Citation `\@citeb' on page \thepage \space undefined}}
        {\csname b@\@citeb\endcsname}}}{#1}}

\newif\if@cghi
\def\cite{\@cghitrue\@ifnextchar [{\@tempswatrue
        \@citex}{\@tempswafalse\@citex[]}}
\def\citelow{\@cghifalse\@ifnextchar [{\@tempswatrue
        \@citex}{\@tempswafalse\@citex[]}}
\def\@cite#1#2{{$\null^{#1}$\if@tempswa\typeout
        {IJCGA warning: optional citation argument 
        ignored: `#2'} \fi}}

 1
 1
 1

\font\ninerm=cmr9

\newcommand{\zp}[3]{Z.\ Phys.\ {\bf C#1} (19#2) #3}
\newcommand{\pl}[3]{Phys.\ Lett.\ {\bf B#1} (19#2) #3}
\newcommand{\plold}[3]{Phys.\ Lett.\ {\bf #1B} (19#2) #3}
\newcommand{\np}[3]{Nucl.\ Phys.\ {\bf B#1} (19#2) #3}
\newcommand{\prd}[3]{Phys.\ Rev.\ {\bf D#1} (19#2) #3}
\newcommand{\prl}[3]{Phys.\ Rev.\ Lett.\ {\bf #1} (19#2) #3}

\newcommand{\md}{\mbox{d}}
\def\simgt{\rlap{\lower 3.5 pt \hbox{$\mathchar \sim$}} \raise 1pt \hbox {$>$}}
\def\simlt{\rlap{\lower 3.5 pt \hbox{$\mathchar \sim$}} \raise 1pt \hbox {$<$}}

\newcommand{\beq}{\begin{equation}}
\newcommand{\eeq}{\end{equation}}
\newcommand{\bea}{\begin{eqnarray}}
\newcommand{\eea}{\end{eqnarray}}

\newcommand{\alps}{\mbox{$\alpha_{\mbox{\scriptsize s}}$}}

\newcommand{\ueight}{\mbox{$\underline{8}$}}

\newcommand{\beqa}{\begin{eqnarray}}
\newcommand{\eeqa}{\end{eqnarray}}
\newcommand{\lqcd}{\mbox{$\Lambda_{\mbox{\scriptsize QCD}}$}}
\begin{document}

\begin{flushright}
RAL-P-97-007\\
July 1997
\end{flushright}

\vspace*{1.5cm}

\boldmath 
\centerline{\normalsize\bf INCLUSIVE $J/\psi$ PHOTOPRODUCTION AT 
    HERA\footnote{Talk presented at the Ringberg Workshop 
    ``New Trends in HERA Physics'', 
    Castle Ringberg, Tegernsee, Germany, 25-30 May
    1997, to appear in the proceedings.}} 
\unboldmath

\vfill
\vspace*{0.6cm}
\centerline{\footnotesize MICHAEL KR\"AMER}
\baselineskip=13pt
\centerline{\footnotesize\it Rutherford Appleton Laboratory}
\baselineskip=12pt
\centerline{\footnotesize\it Chilton, OX11 0QX, UK}
\centerline{\footnotesize E-mail: Michael.Kraemer@rl.ac.uk}

\vspace*{0.9cm} 

\abstracts{I discuss the impact of color-octet contributions to
  inclusive $J/\psi$ photoproduction at \mbox{HERA}.  Emphasis is put
  on resolved photon processes and on $J/\psi$ polarization, which
  will be experimentally accessible at \mbox{HERA} in the near future.
  Both analyses provide a powerful test of the NRQCD factorization
  approach to charmonium production.}
 
\normalsize\baselineskip=15pt
\setcounter{footnote}{0}
\renewcommand{\thefootnote}{\alph{footnote}}

\section{Introduction}
The production of heavy quarkonium states in high-energy collisions
provides an important tool to study the interplay between perturbative
and non-perturbative QCD dynamics. While the creation of heavy quarks
in a hard scattering process can be calculated in perturbative QCD,
the subsequent transition to a physical bound state introduces
non-perturbative aspects. A rigorous framework for treating quarkonium
production and decays has been developed only recently.\cite{BBL95}
The factorization approach is based on the use of non-relativistic QCD
(NRQCD)\cite{CLP} to separate the short-distance parts from the
long-distance matrix elements and explicitly takes into account the
complete structure of the quarkonium Fock space. According to this
formalism, the inclusive cross section for the production of a
quarkonium state $H$ can be expressed as a sum of terms, each of which
factors into a short-distance coefficient and a long-distance matrix
element:
\beq\label{eq_fac} 
\md\sigma(H + X) = \sum_n \md\hat{\sigma}(Q\overline{Q}\, [n] + X)\, 
\langle {\cal{O}}^{H}\,[n]\rangle
\eeq 
Here, $\md\hat{\sigma}$ denotes the short-distance cross section for
producing an on-shell $Q\overline{Q}$-pair in a colour, spin and
angular-momentum state labelled by $n$.  The universal NRQCD matrix
elements $\langle {\cal{O}}^{H} \, [n] \rangle$ give the probability
for a $Q\overline{Q}$-pair in the state $n$ to form the quarkonium
state $H$. The relative importance of the various terms in
(\ref{eq_fac}) can be estimated by using NRQCD velocity scaling
rules.\cite{LMNMH92} For $v\to 0$ ($v$ being the average velocity of
the heavy quark in the quarkonium rest frame) each of the NRQCD matrix
elements scales with a definite power of $v$ and the general
expression (\ref{eq_fac}) can be organized into an expansion in powers
of the heavy quark velocity.

The general factorization formula (\ref{eq_fac}) is believed to hold
also for quarkonium production in hadron-hadron or photon-hadron
collisions.  In the case of inclusive open heavy flavour production it
has been shown that the hadronic cross section factorizes into a
partonic hard scattering cross section multiplied by light quark and
gluon parton densities, with higher-twist corrections being suppressed
by powers $\lqcd/m_Q$.\cite{CSS86} The argument involves averaging
over a sufficiently large range of the heavy quark transverse momentum
$p_T$ and uses the fact that the main contribution to the total
inclusive cross section comes from $p_T\sim m_Q \gg \lqcd$.
Unfortunately, the situation is more involved for quarkonia production
in hadronic collisions, since the region of small transverse
$Q\overline{Q}$ momentum significantly contributes to the total cross
section. At small transverse momentum however, the intermediate
colour-singlet or colour-octet $Q\overline{Q}$ pair moves parallel with
a remnant jet and might interact before the physical bound state has
been formed.\footnote{Higher-twist corrections to colour-singlet
  $J/\psi$ photoproduction have been analysed recently.\cite{MA97}}
Only in the heavy quark limit, where all scales involved in the bound
state physics are much larger than $\lqcd$, higher-twist terms are
necessarily strongly suppressed.  For charmonium production in
hadronic collisions, it seems however reasonable to restrict the
analysis to the region $p_T \gg \lqcd$, where the general
factorization formula (\ref{eq_fac}) should safely be applicable.
Excluding the small $p_T$ region is also necessary to guarantee
perturbative stability in higher-order QCD calculations of $J/\psi$
photoproduction.\cite{KZSZ94,MK95}

The NRQCD formalism implies that colour-octet processes, in which the
heavy-quark antiquark pair is produced at short distances in a
colour-octet state and subsequently evolves nonperturbatively into a
physical quarkonium, must contribute to the cross section. As
discussed extensively in the literature\cite{BF95,tev}, colour-octet
contributions appear as the most plausible explanation of the large
direct $\psi$ production cross section observed at the
Tevatron\cite{cdf}.  The NRQCD approach is certainly the correct
theory for quarkonium production in the heavy quark limit. It is
however not clear whether the charm quark mass is sufficiently large
to allow for a reliable expansion in the heavy quark velocity.  In
order to establish the phenomenological significance of the
colour-octet mechanism and the universality of the NRQCD matrix
elements for charmonium production it is thus necessary to identify
colour-octet contributions in different production processes. The
analysis of $J/\psi$ photoproduction at HERA appears to be a very
powerful tool to constrain the colour-octet matrix elements and to
test the picture of quarkonium production as developed in the context
of the NRQCD factorization approach.

The production of $J/\psi$ particles in high energy $ep$ collisions at
\mbox{HERA} is dominated by photoproduction events where the electron
is scattered by a small angle producing photons of almost zero
virtuality. The measurements at \mbox{HERA} provide information on the
dynamics of $J/\psi$ photoproduction in a wide kinematical region,
$30~\mbox{GeV} \; \simlt \; \sqrt{s\hphantom{tk}}\!\!\!\!\!  _{\gamma
  p}\;\simlt\; 200~\mbox{GeV}$, corresponding to initial photon
energies in a fixed-target experiment of $450~\mbox{GeV} \; \simlt \;
E_\gamma \; \simlt \;$ $20,000~\mbox{GeV}$.  The production of
$J/\psi$ particles in photon-proton collisions proceeds predominantly
through photon-gluon fusion, where the photon interacts directly with
the gluon from the proton. Besides the direct photoproduction channel,
$J/\psi$ production at \mbox{HERA} can also take place via resolved
photon contributions where the photon behaves as a source of partons
which interact with the partons in the proton. Resolved processes are
expected to contribute significantly to the lower endpoint of the
$J/\psi$ energy spectrum and might be probed at \mbox{HERA} for the
first time in the near future.\footnote{Elastic mechanisms only
  contribute to the region of small $J/\psi$ transverse momentum and
  can be suppressed by an appropriate cut in $p_T$.\cite{elastic}}

\section{Direct Photon Contributions}
For $J/\psi$ production and at leading order in the velocity
expansion, the general factorization formula (\ref{eq_fac}) reduces to
the standard expression of the colour-singlet model\cite{CS}. The
short-distance cross section for $J/\psi$ photoproduction through
direct photons is given by the subprocess
\begin{equation}\label{eq_cs}
\gamma + g \to c\bar{c}\, [^3\!S_1,\underline{1}] + g
\end{equation}
with $c\bar{c}$ in a colour-singlet state (denoted by
\mbox{$\underline{1}$}), with zero relative velocity, and
spin/angular-mo\-men\-tum quantum numbers $^{2S+1}L_J = {}^3\!S_1$.
Relativistic corrections to the colour-singlet channel due to the
motion of the charm quarks in the $J/\psi$ bound state enhance the
small-$p_t$ region, but can be neglected for $p_T\;\simgt\;
1$~GeV.\cite{REL} The calculation of the higher-order perturbative QCD
corrections to the short-distance cross section (\ref{eq_cs}) has been
performed recently.\cite{KZSZ94,MK95} Inclusion of the NLO corrections
reduces the scale dependence of the theoretical prediction and
increases the cross section significantly, depending in detail on the
$\gamma p$ energy and the choice of parameters.\cite{MK95} Details of
the calculation and a comprehensive analysis of total cross sections
and differential distributions for the energy range of the
fixed-target experiments and for $J/\psi$ photoproduction at
\mbox{HERA} can be found elsewhere.\cite{MK95}

The leading colour-octet configurations which contribute to $J/\psi$
photoproduction at $p_T \neq 0$ are produced through the
subprocesses\cite{CK96}
\begin{eqnarray}\label{eq_co}
\gamma + g &\! \to \!& c\bar{c}\, 
[{}^1\!S_{0},{}^3\!S_{1},{}^3\!P_{0,1,2},\underline{8}] + g \nonumber\\
\gamma + q(\bar{q}) &\! \to \!& c\bar{c}\, 
[{}^1\!S_{0},{}^3\!S_{1},{}^3\!P_{0,1,2},\underline{8}] + q(\bar{q}) 
\end{eqnarray}
The transition of the colour-octet $c\bar{c} \,
[{}^{2S+1}L_{J},\mbox{$\underline{8}$}]$ pairs into a physical
$J/\psi$ state through the emission of non-perturbative gluons is
described by the long-distance matrix elements $\langle
{\cal{O}}^{J/\psi} \, [{}^{2S+1}L_{J},\mbox{$\underline{8}$}]
\rangle$.  They have to be obtained from lattice si\-mu\-la\-ti\-ons
or measured directly in some production process.  According to the
velocity scaling rules of NRQCD, the colour-octet matrix elements
associated with $S$-wave quarkonia should be suppressed by a factor of
$v^4 \sim 10^{-2}$ compared to the leading colour-singlet matrix
element.  However, explicit calculation of the processes (\ref{eq_co})
shows that the short-distance factors of the
$[{}^{1}\!S_{0},\mbox{$\underline{8}$}]$ and
$[{}^{3}\!P_{0,2},\mbox{$\underline{8}$}]$ channels are strongly
enhanced as compared to the colour-singlet term and compensate the
${\cal{O}}(10^{-2})$ suppression of the corresponding non-perturbative
matrix elements. This short-distance enhancement is due to $t$-channel
gluon exchange which contributes to the
$[{}^{1}\!S_{0},{}^{3}\!P_{0,2},\mbox{$\underline{8}$}]$ cross
sections, but not to the leading-order colour-singlet channel.

\section{Resolved photon contributions}
Inclusive $J/\psi$ photoproduction can also take place via resolved
photon interactions, where the photon first splits into a flux of
light quarks and gluons which then may fuse with a gluon or light
quark from the proton to form a $c\bar{c}$ pair. This process has been
extensively analyzed in the past in the framework of the colour-singlet
model, where the cross section is given by gluon-gluon fusion into a
${}^3\!S_1$ colour-singlet $c\bar c$ pair, which subsequently
hadronizes into a $J/\psi$:
\begin{equation}\label{eq_csres}
g + g \to c\bar{c}\, [{}^3\!S_1,\underline{1}] + g
\end{equation}
The process (\ref{eq_csres}) contributes to the overall cross section
only marginally except near the lower endpoint of the $J/\psi$
energy spectrum.\cite{JST92}

Within the NRQCD factorization approach, however, more resolved-photon
channels have to be considered, where $J/\psi$ production proceeds via
colour-octet $c\bar{c}$ pairs.\cite{CK97} The leading colour-octet
contributions are:
\begin{eqnarray}\label{eq_cores}
g + g &\! \to \!& c\bar{c}\, [{}^1\!S_{0},{}^3\!S_{1},{}^3\!P_{0,1,2},
\underline{8}] + g \nonumber \\
g + q(\bar{q}) &\! \to \!& c\bar{c}\, [{}^1\!S_{0},{}^3\!S_{1},
{}^3\!P_{0,1,2},
\underline{8}] + q(\bar{q}) 
\nonumber \\
q + \bar{q} &\! \to \!& c\bar{c}\, [{}^1\!S_{0},{}^3\!S_{1},
{}^3\!P_{0,1,2},
\underline{8}] + g
\end{eqnarray}
They are in every respect analogous to the ones which have been argued
to strongly increase the $J/\psi$ production cross section in $p\bar
p$ collisions at the Tevatron.\cite{BF95,tev} Colour-octet
contributions to resolved $J/\psi$ photoproduction thus provide an
important direct test of the NRQCD explanation of the Tevatron data.

\boldmath 
\section{The $J/\psi$ Energy Distribution}
\begin{figure}[ht]
  \vspace{-2.3cm} \epsfysize=16cm \epsfxsize=12cm
  \centerline{\epsffile{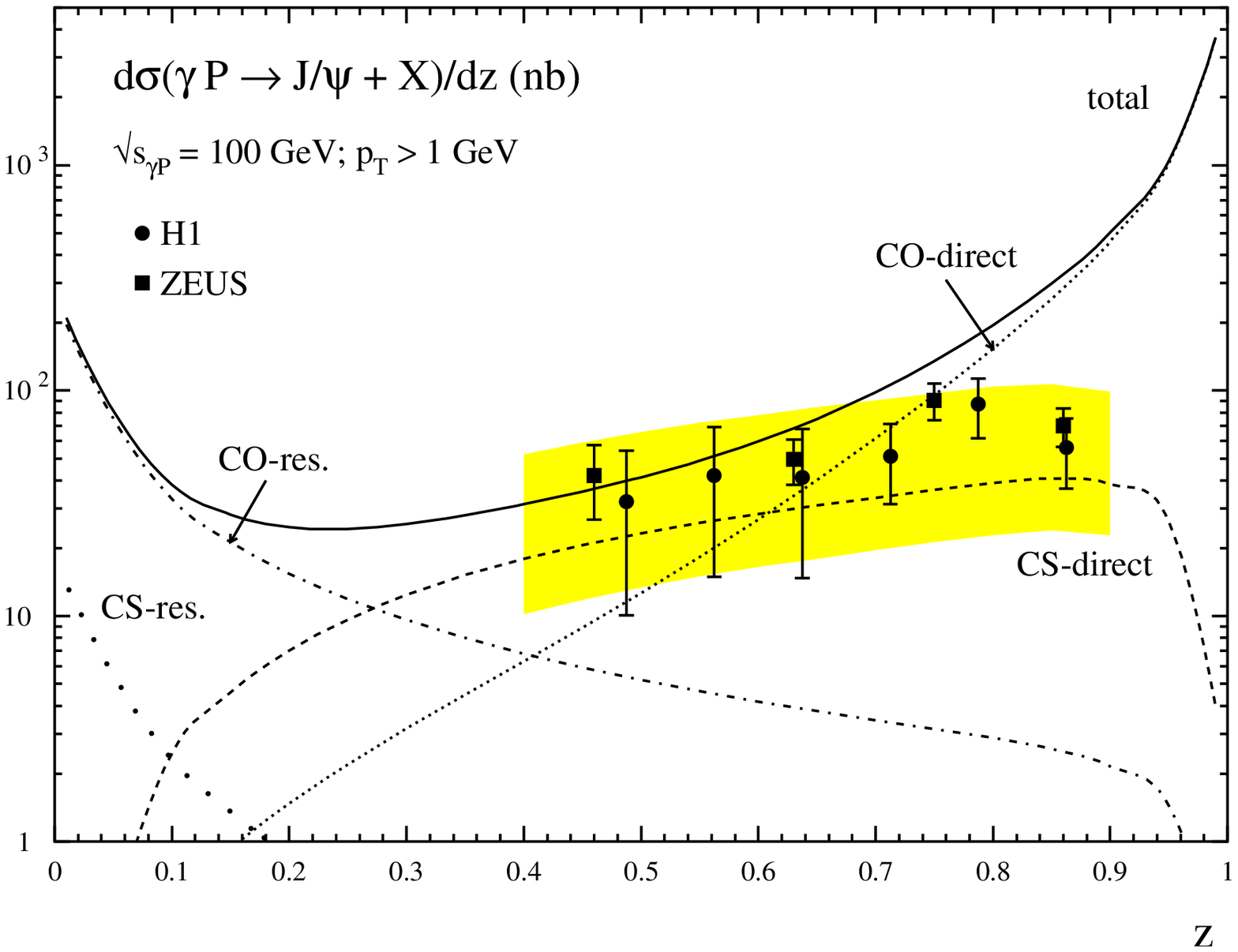}} \vspace*{-3cm}
  \fcaption{Colour-singlet (CS) and colour-octet (CO) contributions to
    the $J/\psi$ energy distribution $\md\sigma/\md{}z$ at the
    photon-proton centre of mass energy
    $\sqrt{s\hphantom{tk}}\!\!\!\!\!  _{\gamma p}\,\, = 100$~GeV
    integrated in the range $p_T \ge 1$~GeV, compared to experimental
    data\cite{H1,ZEUS}. Both direct and resolved photon processes are
    included. Parameters: $m_c = 1.5$~GeV,
    renormalization/factorization scale $\mu = 2 m_c$, GRV parton
    distribution functions\cite{pdf} with $\Lambda^{(4)} = 200$~MeV.
    The shaded band reflects the uncertainty in the normalization of
    the direct colour-singlet contributions due to variation of $m_c$
    and $\alps$, as described in the text.}
\label{fig1}
\end{figure}

\unboldmath 
The $J/\psi$ energy distribution $\mbox{d}\sigma/\mbox{d}{}z$ offers
unique possibilities to study the relative importance of the different
mechanisms that contribute to $J/\psi$ photoproduction.  The scaling
variable $z$ is defined by $z = {p\cdot k_\psi}\, / \, {p\cdot
  k_\gamma}$, $p, k_{\psi,\gamma}$ being the momenta of the proton and
$J/\psi$, $\gamma$ particles, respectively. In the proton rest frame,
$z$ is the ratio of $J/\psi$ to $\gamma$ energy, $z=E_\psi/E_\gamma$.
Adopting NRQCD matrix elements consistent with those extracted from
the fits to prompt $J/\psi$ data at the Tevatron\cite{tev} (see Table
\ref{table1}), one finds significant colour-octet contributions near the
upper and lower endpoint of the $J/\psi$ energy spectrum.
\begin{table}[thb]
\renewcommand{\arraystretch}{1.1}
\begin{center}
\tcaption{\label{table1}
\small Values of the NRQCD matrix elements used in the numerical
 analysis, with the velocity and mass scaling.}
\vspace*{3mm}
\begin{tabular}{|c|c|c|}
\hline
$\langle {\cal{O}}^{J/\psi}\,[^{3}\!S_{1},\underline{1}]\rangle
\hphantom{/m_c^2}$ &$1.16$ GeV$^3$ & $\quad m_c^3 v^3$ \\
$\langle {\cal{O}}^{J/\psi}\,[^{1}\!S_{0},\underline{8}]\rangle
\hphantom{/m_c^2}$ &$10^{-2}$ GeV$^3$ & $\quad m_c^3 v^7$ \\
$\langle {\cal{O}}^{J/\psi}\,[^{3}\!S_{1},\underline{8}]\rangle
\hphantom{/m_c^2}$ &$10^{-2}$ GeV$^3$ & $\quad m_c^3 v^7$ \\
$\langle {\cal{O}}^{J/\psi}\,[{}^{3}\!P_{0},\underline{8}]\rangle 
/ m_c^2$
& $10^{-2}$ GeV$^3$ & $\quad m_c^3 v^7$ \\
\hline
\end{tabular}
\end{center}
\renewcommand{\arraystretch}{1}
\end{table}
This is demonstrated in Fig.~\ref{fig1}, where I plot the
colour-singlet and colour-octet contributions to
$\mbox{d}\sigma/\mbox{d}{}z$ at a typical \mbox{HERA} energy of
$\sqrt{s\hphantom{tk}} \!\!\!\!\!  _{\gamma p}\,\, = 100$~GeV in the
restricted range $p_\perp \ge 1$~GeV via direct and resolved photon
processes.\footnote{Note that a cut in the $J/\psi$ transverse
  momentum, $p_t\;\simgt\; 1$~GeV, is sufficient to exclude the
  elastic contributions; no additional cut in $z$ is required.} The
colour-octet processes are predicted to exceed the colour-singlet ones
by more than an order of magnitude for $z\;\simgt\;0.6$ and
$z\;\simlt\; 0.2$. Resolved contributions are negligible for
$z\;\simgt\; 0.3$, but colour-octet terms are expected to become
significant at the lower end of the energy
spectrum.\footnote{Similarly, colour-octet processes enhance the
  low-$z$ region in $J/\psi$ photoproduction at large $p_T$ via
  fragmentation mechanisms.\cite{KK97}} The theoretical predictions
are compared to experimental results obtained by the
\mbox{H1}\cite{H1} and \mbox{ZEUS}\cite{ZEUS} collaborations at
\mbox{HERA}. From Fig.~\ref{fig1} one can conclude that the large
colour-octet enhancement in the region $z\;\simgt\;0.6$ is not
supported by the experimental data and that the $J/\psi$ energy
spectrum is adequately described by the colour-singlet channel.
Taking into account the large uncertainty of the cross section
normalization due to the choice for the charm quark mass and the QCD
coupling, the current experimental data on the $J/\psi$ energy
distribution can be accounted for by the colour-singlet channel alone.
This can be judged by comparing the experimental data with the shaded
band in Fig.~\ref{fig1} which indicates how much the colour-singlet
cross section is altered if $m_c$ and $\Lambda^{(4)}$ vary in the
range $1.35~\mbox{GeV} < m_c < 1.55~\mbox{GeV}$ and $200~\mbox{MeV}<
\Lambda^{(4)} < 300~\mbox{MeV}$.\footnote{To study the $\alps$
  dependence of the cross section, I use consistently adjusted sets of
  parton densities.\cite{pdf}}

\begin{figure}[th]
   \vspace{-2.3cm}
   \epsfysize=16cm
   \epsfxsize=12cm
   \centerline{\epsffile{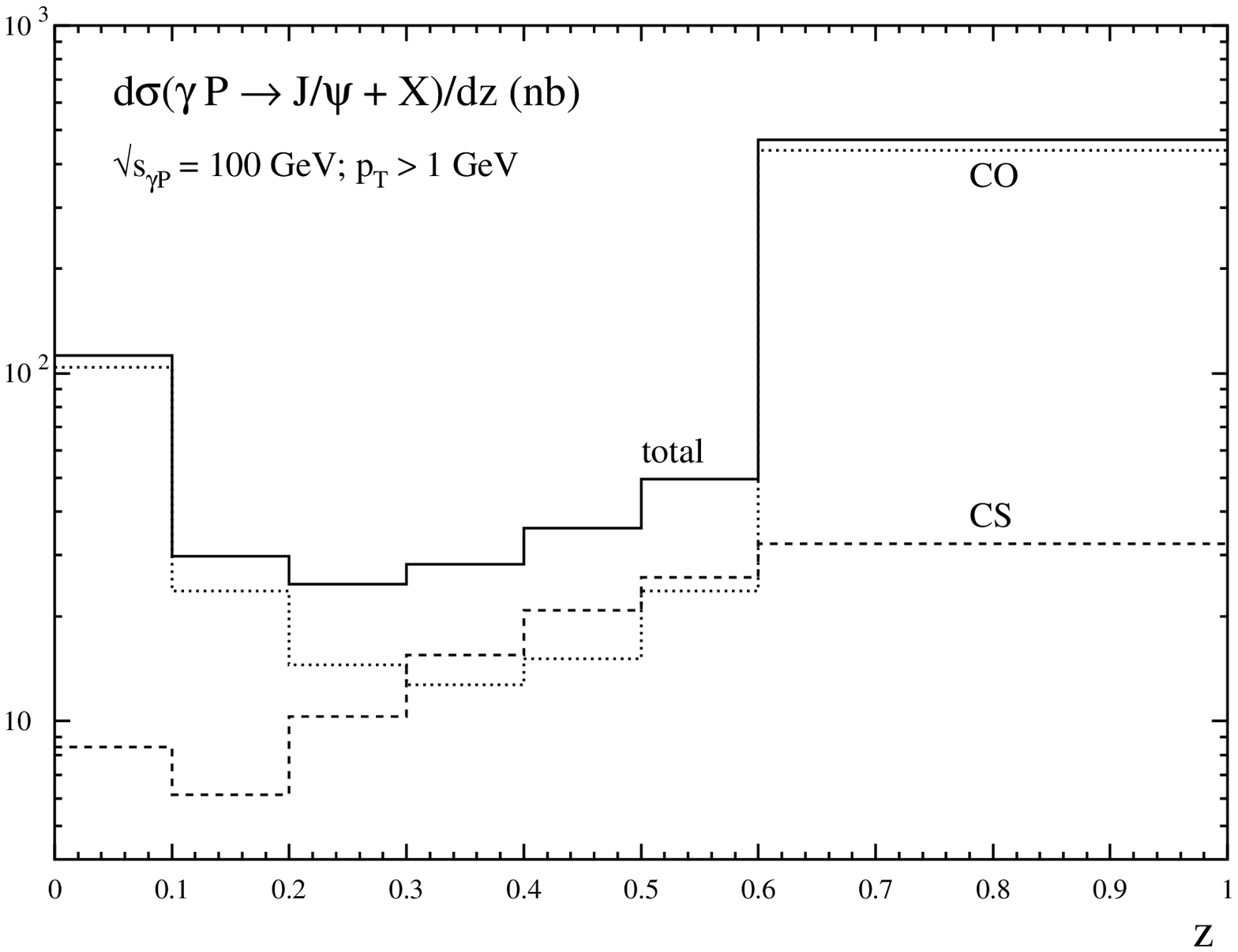}}
   \vspace*{-3cm}
   \fcaption{Same as Fig.~\ref{fig1}, averaged over large intervals
     in $z$.}
\label{fig2}
\end{figure}

One should be careful to interpret the discrepancy between the
theoretical predictions for the colour-octet contributions and the
\mbox{HERA} data shown in Fig.~\ref{fig1} as a failure of the
factorization approach. As pointed out recently, the NRQCD expansion
breaks down close to the phase space boundary $z\to 1$, and no
prediction can be made for $z\;\simgt\;0.75$ without resumming large
corrections in the velocity expansion.\cite{BRW97} Therefore, the
shape of $\mbox{d}\sigma/ \mbox{d}{}z$ near the endpoint region cannot
be used to access the importance of colour-octet contributions.
Nevertheless, the analysis of the the $J/\psi$ energy distribution
provides a crucial constraint on the colour-octet matrix elements if
one averages over a sufficiently large interval $\Delta z \gg v^2 \sim
0.25$ containing the endpoint $z=1$, as shown in Fig.~\ref{fig2}.
Colour-octet mechanisms should contribute significantly to the region
$0.6\;\simlt\;z\le 1$ and exceed the colour-singlet cross section by
more than an order of magnitude. While the overall size of the cross
section is subject to large uncertainties due to $m_c$ and $\alps$,
the distinctive modification of the $J/\psi$ energy spectrum by
colour-octet processes in the low and large $z$ region should clearly
be visible in the experimental data. Moreover, the low-$z$ region of
the energy spectrum is not expected to be sensitive to higher order
terms in the velocity expansion and can thus reliably be predicted
within leading-twist NRQCD.

The results at \mbox{HERA} seem to indicate that the values of the
colour-octet matrix elements $\langle {\cal{O}}^{J/\psi} \,
[{}^{1}\!S_{0},\ueight]\rangle$ and $ \langle {\cal{O}}^{J/\psi} \,
[{}^{3}\!P_{0},\ueight] \rangle$ are considerably smaller than
suggested by the fits to the CDF data at the Tevatron. This does
however not necessarily imply the non-universality of the NRQCD matrix
elements for $J/\psi$ production. The determination of the $\langle
{\cal{O}}^{J/\psi} \, [{}^{1}\!S_{0},\ueight]\rangle$ and $ \langle
{\cal{O}}^{J/\psi} \, [{}^{3}\!P_{0},\ueight] \rangle$ matrix elements
from the Tevatron data is very sensitive to all effects that modify
the shape of the $J/\psi$ transverse momentum distribution at
$p_t\;\simlt\; 5$~GeV and might yield much too large
values.\cite{tev2} The theoretical uncertainties include the small-$x$
behaviour of the gluon distribution, the evolution of the strong
coupling, next-to-leading order contributions, intrinsic transverse
momentum of the partons in the proton, and also systematic effects
inherent in NRQCD, such as the inaccurate treatment of the energy
conservation in the hadronization of the colour-octet $c\bar{c}$
pairs. With higher statistics data at \mbox{HERA} it will be possible
to extract more detailed information on the colour-octet matrix
elements, in particular from the analysis of the $J/\psi$ energy
distribution.

\boldmath 
\section{$J/\psi$ Polarization}
\begin{figure}[th]
  \vspace{-2.3cm} 
  \epsfysize=16cm 
  \epsfxsize=12cm
  \centerline{\epsffile{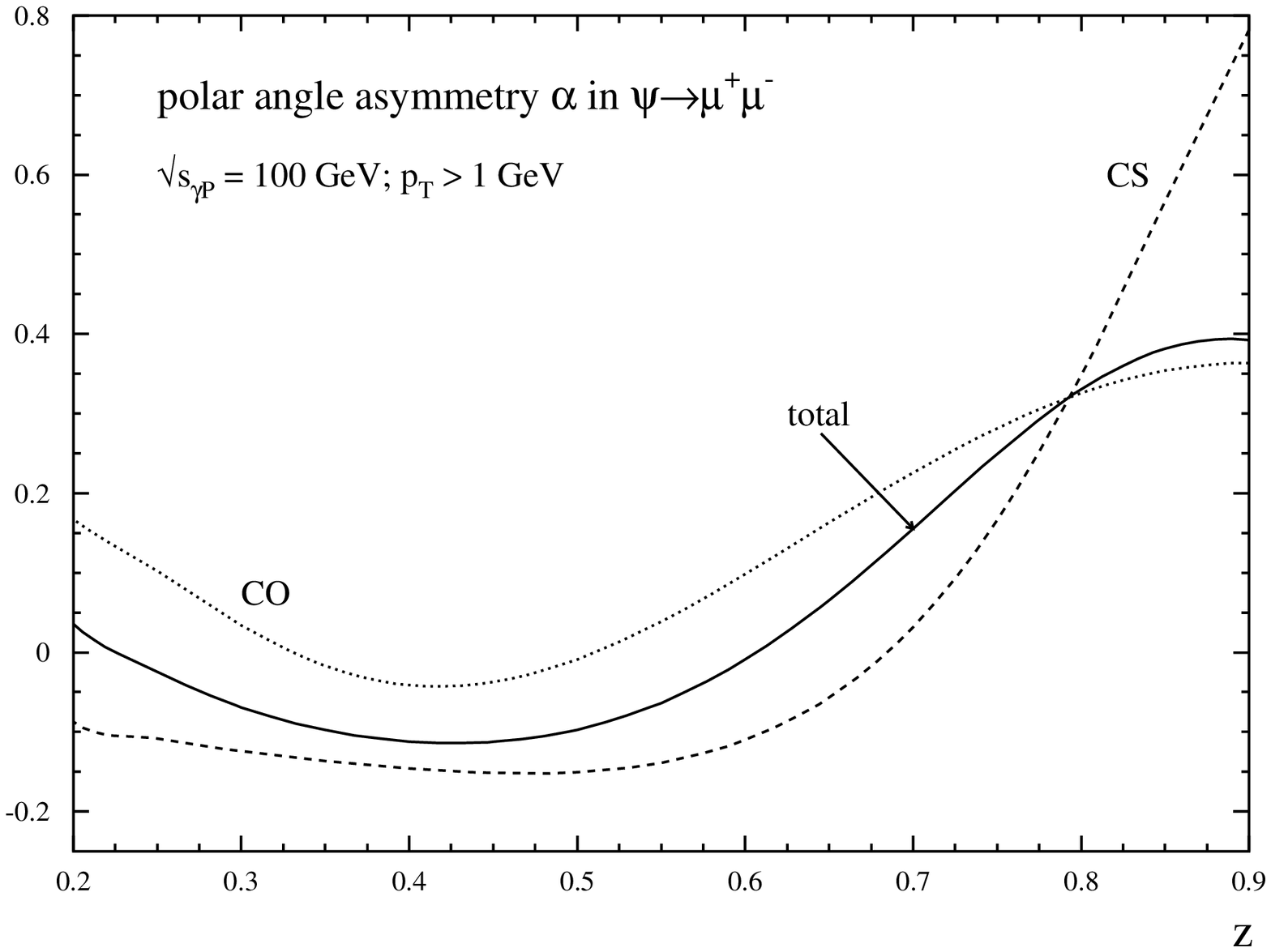}} 
  \vspace*{-3cm}
  \fcaption{Colour-singlet (CS) and colour-octet (CO) contributions to
    the polar angle asymmetry $\alpha$ in (\ref{eq_ang}) at the
    photon-proton centre of mass energy
    $\sqrt{s\hphantom{tk}}\!\!\!\!\!  _{\gamma p}\,\, = 100$~GeV
    integrated in the range $p_T \ge 1$~GeV. Both direct and resolved
    photon processes are included.}
\label{fig3}
\end{figure}

\unboldmath
The NRQCD factorization approach allows for unambiguous predictions of
the quarkonium polarization.\cite{pol} Transverse polarization of
$J/\psi$ and $\psi'$ particles, produced directly at large transverse
momentum in $p\bar{p}$ collisions at the Fermilab Tevatron, has in
fact emerged as the most prominent test of colour-octet contributions
and spin symmetry in charmonium production.\cite{CW95} Polarization
signatures in $J/\psi$ photoproduction have been analysed in detail in
the past within the colour-singlet model.\cite{polcs} Recently, a
comprehensive calculation has been performed in the context of the
NRQCD factorization approach, including colour-singlet and colour-octet
contributions to both direct as well as resolved photon
processes.\cite{BKV97} Results are available for the most general
decay angular distribution in any given reference frame. Here, I only
consider the polar-angle distribution in the $s$-channel helicity
frame, in which the polarization axis in the $J/\psi$ rest frame is
defined as the direction of the $J/\psi$ three-momentum in the
photon-proton cms frame. The polar angular distribution in the decay
$J/\psi\to l^+l^-$ is given by
\begin{equation}\label{eq_ang}
\frac{\md\Gamma}{\md\cos\theta}\propto 1+\alpha\,\cos^2\theta,
\end{equation}
with $\theta$ the angle between the lepton three-momentum in the
$J/\psi$ rest frame and the polarization axis.  Fig.~\ref{fig3}
displays colour-singlet and colour-octet contributions to the polar
asymmetry $\alpha$, defined in (\ref{eq_ang}), as function of the
$J/\psi$ energy variable $z$.
It is worth pointing out that the asymmetry predictions involve ratios
of cross sections and are thus to a large extent insensitive to
uncertainties in the charm quark mass, the QCD coupling or parton
distribution functions. Therefore, polarization measurements are a
rather clean test of the underlying quarkonium production mechanisms
and will allow to access the relative importance of colour-singlet and
colour-octet contributions.

\section{Conclusions}
I have discussed colour-singlet and colour-octet contributions to
$J/\psi$ photoproduction at \mbox{HERA}. The high-statistics data to
be expected in the near future will allow the analysis of resolved
photon processes and $J/\psi$ polarization to further test the
phenomenological significance of colour-octet contributions. Various
other channels and final states can be accessed in the future at
\mbox{HERA}, like photoproduction of $\psi'$, $\Upsilon$ and $\chi_c$
particles\cite{CK97,MA96}, associated $J/\psi + \gamma$
production\cite{CK97,TM97,CGK97}, fragmentation contributions at large
$p_T$\cite{GRS95,KK97} as well as deep inelastic $J/\psi$
production\cite{SF97}. The future analyses of inclusive quarkonium
production at \mbox{HERA} offer unique possibilities to test the NRQCD
factorization approach to charmonium production.

\section{Acknowledgements}
I would like to thank Martin Beneke, Matteo Cacciari and Mikko
V\"anttinen for their collaboration on quarkonium production.

\end{document}